\documentstyle[aps]{revtex} 
\begin{document}
\draft
\title{Exact Solutions with w - modes}
\author{Mustapha Ishak, Luke Chamandy, Nicholas Neary  and Kayll Lake\cite{email}}
\address{Department of Physics, Queen's University, Kingston, Ontario, Canada, K7L 3N6 }
\date{\today}
\maketitle
\begin{abstract}
An explicit necessary condition for the internal trapping of null geodesics, along with the occurrence of resonance scattering of axial gravitational waves, is proposed for static spherically symmetric perfect fluid solutions of Einstein's equations. Some exact inhomogeneous solutions which exhibit this trapping are given with special attention to boundary conditions and the physical acceptability of the space times. In terms of the tenuity ($\alpha = R/M$ at the boundary) all the examples given here lie in the narrow range $2.8 < \alpha < 2.9$. The tenuity can be raised to more interesting values by the addition of an envelope without altering the trapping.

\end{abstract} 
\section{Introduction}It is now well known that sufficiently compact stars (polytropic or uniform density) can support the internal trapping of null geodesics and the "w-modes" found by Chandrasekhar and Ferrari (1991)\cite{CHANDRAFER}, \cite{vishu}. These modes exist both for axial and polar perturbations, though the axial ones have been studied more thoroughly. The w-modes in general have no Newtonian counterparts\cite{KOJI} since they are predominantly modes of the spacetime. In the polar case they couple weakly to the fluid while in the axial case there is no coupling at all. Recent numerical studies of these w-modes have involved the effect of the equation of state \cite{BBFER} and their excitation \cite{ap},\cite{TOMSAIMAE}. Whereas the role that these w-modes may play in real astrophysical processes remains open to much further investigation, it is fair to say that little is actually known about the behavior of the governing potential of the wave equation in exact solutions of Einstein's equations. Such knowledge is important since it is both a route to the physical understanding of relativistic phenomena and a check on numerical procedures. The purpose of this paper is to explore necessary conditions for the internal trapping of null geodesics and the existence of w-modes (when the centrifugal part of the potential dominates) in physically acceptable exact isolated static spherically symmetric perfect fluid solutions of Einstein's equations. We are able to exhibit physically acceptable exact solutions which have trapping and which could support w-modes.
\section{Review of Perfect Fluids}
Any metric is an ``exact" solution to Einstein's equations. However, the consequent energy-momentum tensor is almost never of any interest. What is of interest are solutions which might have some contact with reality. Recently \cite{del} a collection of exact isolated static spherically symmetric perfect fluid solutions have been subjected to the following elementary criteria for physical acceptability:
\begin{itemize}
\item[1.]{Isotropy of the pressure ($p$).}
\item[2.]{Regularity of the origin by way of the scalars polynomial in the Riemann tensor \cite{LakeMus}, \cite{ell}, \cite{schmidt}.}
\item[3.]{Positive definiteness of both $p$ and energy density ($\rho$)  
at the origin.}
\item[4.]{Isolation by way of the requirement that the pressure reduce to zero at some finite 
boundary radius $r_{\Sigma} > 0$.}
\item[5.]{Monotonicity of both $p$ and $\rho$ to the boundary.}
\item[6.]{Subluminal adiabatic sound speed ($\textrm{v}_s^2 = \frac{dp}{d\rho}
< 1$) \cite{C-B}.}
\end{itemize}
Perhaps not surprisingly, only about 10 \% of the solutions pass these elementary tests. In what follows we take the view that solutions worthy of further consideration must pass all the applicable tests in at least some region \cite{schwint}. We also take the view that an analytic solution of Einstein's equations can be expected to approximate only a region of a realistic configuration. That is, an analytic solution could have an interior causal limit ($\textrm{v}_s^2 =1$), a circumstance which precludes standard stability arguments \cite{stability}, and yet provide an adequate approximation for a region of a realistic configuration.

We begin by setting the notation. The line element in conventional form is (e.g., \cite{MTW}) \cite{units}
\begin{equation}
ds^2=\frac{dr^2}{1-\frac{2m(r)}{r}}+r^2(d\theta^2+sin(\theta)^2d\phi^2)-e^{2\Phi(r)}dt^2 \label{standard}
\end{equation}
with the coordinates comoving in the sense that the fluid streamlines are given by $u^{a}=e^{- \Phi(r)}\delta^{a}_{t}$. In terms of the functions $\Phi(r)$ and $m(r)$ the regularity conditions reduce to 
\begin{equation}
\Phi'(0)=m(0)=m'(0)=0, \label{reg}
\end{equation}
with $' \equiv d/dr$ and $\Phi(0)$ a constant fixed by the scale of $t$. Next, in terms of the perfect fluid decomposition ($T^{a}_{b}=(\rho(r)+p(r))u^{a}u_{b}+p(r)\delta^{a}_{b}$), solving for $\Phi'(r)$ from the $r$-component of the conservation equations and Einstein's equations  ($\nabla_{a}T^{a}_{r}=0$ and $G^{r}_{r}-8\pi p(r)=0$) we obtain the Tolman \cite{tolman1} -Oppenheimer-Volkoff \cite{ov} (TOV) equation
\begin{equation}
\Phi'(r)=\frac{-p'(r)}{\rho(r)+p(r)}=\frac{m(r)+4\pi p(r)r^3}{r(r-2m(r))}, \label{tov}
\end{equation}
where, from the $t$ component of the Einstein equations ($G^{t}_{t}=-8\pi \rho(r)$),
\begin{equation}
4\pi \rho(r)=\frac{m'(r)}{r^2}. \label{rho}
\end{equation}
From the TOV equation (here taken to be the right hand members of  (\ref{tov})) we observe that $p(r)$ is maximal at $r=0$. Moreover, if there is an equation of state ($p(\rho)$) then either $\rho$ is maximal wrt $r$ at $r=0$ or $p$ is maximal wrt $\rho$ at $r=0$. Despite that fact that the TOV equation has been known for over sixty years, only recently \cite{br} has its mathematical structure been fully appreciated. For example, we now know that for $p(r)>0$ there exists a unique global solution for every $0<p(0)<\infty$. It is not difficult to find ``solutions" of the TOV equation. For example, $m(r)$ can be chosen in such a way that (\ref{tov}) yields a solution (with $\rho(r)$ following from (\ref{rho})). The simplest choice is clearly $m \propto r^3$ but this leads us back to the Schwarzschild interior solution. 
The metric (\ref{standard}) contains two functions, $m(r)$ and $\Phi(r)$, related by (\ref{tov}). The first represents the gravitational energy (effective gravitational mass) (e.g., \cite{hayward}). The second is, in the weak field limit $r \gg 2m(r)$, the Newtonian potential. This interpretation offers no insight into the meaning of $\Phi(r)$ within Einstein's theory, and is a good point to begin our discussion.
\section{Null Geodesic Limit}
We start with the ``centrifugal" part of the potential \cite{sonego} for non-radial odd parity perturbations. This governs the evolution of null geodesics. Radial null geodesics of the metric (\ref{standard}) satisfy
\begin{equation}
t=\pm \int \frac{dr}{ e^{\Phi(r)} \sqrt{1-\frac{2m(r)}{r}}}+{\em D},\label{radial}
\end{equation}
with $\theta$, $\phi$, and {\em D} constant. Non-radial null geodesics satisfy $\theta=\pi/2$ (by choice),
\begin{mathletters}
\label{allequations}
\begin{equation}
r^{4}\phi^{\bullet2}=1,\label{nra}
\end{equation}
\begin{equation}
e^{4\Phi(r)}t^{\bullet2}=\frac{1}{b^{2}},\label{nrb}
\end{equation}
and
\begin{equation}
r^{2}r^{\bullet2}=(1-\frac{2m(r)}{r})(\frac{B(r)^{2}}{b^{2}}-1),\label{nrc}
\end{equation}
with
\begin{equation}
B(r) \equiv re^{-\Phi(r)},\label{nrd}
\end{equation}
\end{mathletters}
where $ \bullet \equiv d/d\lambda$ for affine $\lambda$, and $b$ is a constant $>0$, the ``impact parameter". The ``potential" impact parameter $B(r)$ provides, by way of  (\ref{nrd}), an invariant physical interpretation of  $\Phi(r)$.  Null geodesics are restricted by the condition $b \leq B(r)$ \cite{nl}. 

From conditions (\ref{reg}) and the definition (\ref{nrd}) it follows that
\begin{equation}
B(r) \sim \psi_{0}r \label{lim}
\end{equation}
as $r \rightarrow 0$ where $\psi_{0}$ is a physically irrelevant scale factor. (The ratio $B(r)/b$ is invariant to scale changes in $t$.) It follows that the necessary and sufficient condition for the internal trapping of null geodesics (that is the existence of $r_{0}$ such that $r^{\bullet}=0$ and $r^{\bullet \bullet}<0$ at $r_{0}$) is given by 
\begin{mathletters}
\label{allequations}
\begin{equation}
\Phi'(r)>\frac{1}{r} \label {trapa}
\end{equation}
or, from (\ref{tov}),
\begin{equation}
p'(r)<-\frac{\rho(r)+p(r)}{r} \label {trapb}
\end{equation}
which, with an equation of state ($p(\rho)$) can be given as
\begin{equation}
\rho'(r)<-\frac{\rho(r)+p(r)}{v_{s}^{2}r}. \label {trapc}
\end{equation}
\end{mathletters}
From (\ref{trapa}) and the TOV equation it follows that
\begin{equation}
r<3m(r)+4 \pi p(r)r^{3}, \label{tov1}
\end{equation}
a relation which makes the trapping of null geodesics a manifestly relativistic phenomenon \cite{bschw}, \cite{cve}.
\section{Full Potential}
The odd parity (axial) w-modes are non-radial perturbations of the spacetime
which do not couple to the fluid at all. In terms of the frequency $\varpi$ and mode number $l \geq 2$ the governing equation is given by \cite{cvorig}
\begin{equation}
	(\frac{d^2 }{d{r_*}^2}+\varpi^2)Z=V(r_*) Z , \label{wave}
\end{equation}
where $r_*$ is the ``tortoise" coordinate
\begin{equation}
	dr_* = \frac{e^{-\Phi(r)}}{\sqrt{1-\frac{2 m(r)}{r}}}dr. \label{tortoise}
\end{equation}
The potential is conveniently expressed in terms of $r$ and is given by
\begin{equation}
	V(r)=\frac{1}{B(r)^2}(l(l+1)+4\pi r^2(\rho(r)-p(r))-\frac{6 m(r)}{r}). \label{potential}
\end{equation}
A necessary condition for the occurrence of resonance scattering of axial gravitational waves by an isolated distribution of fluid is a local minimum in $V(r)$ within the boundary of the fluid. (If the centrifugal part of the potential ($\frac{1}{B(r)^2}(l(l+1)$) dominates, which is frequently but not always the case (see below), then (\ref{tov1}) provides such a condition.) It is the purpose of this paper to explore the occurences of this minimum in physically acceptable exact solutions. It is the shape of the function $V(r)$ which is of interest, and since the exterior vacuum has a well known local maximum at $r \sim 3.28 M$ (for $l=2$), the boundary conditions associated with the fluid - vacuum interface need careful attention.
\section{Boundary Conditions}
The Darmois-Israel junction conditions demand the continuity of the first and second fundamental forms at a boundary surface. These conditions are well known (e.g., \cite{MusLake}) but are usefully reviewed here. We take the ``interior" metric to be of the form (\ref{standard}). The ``exterior" is the familiar Schwarzschild vacuum (in coordinates ($\texttt{r}\not=r, \theta, \phi, T\not=t$)):
\begin{equation}
ds^2=\frac{d\texttt{r}^2}{1-\frac{2M}{\texttt{r}}}+\texttt{r}^2(d\theta^2+sin(\theta)^2d\phi^2)-(1-\frac{2M}{\texttt{r}})dT^2. \label{vacuum}
\end{equation}
At the fluid interface ($\Sigma$), without loss in generality, we take $\theta$ and $\phi$ continuous (with intrinsic coordinates $\theta, \phi, \tau$, where $\tau$ is the proper time). This gives
\begin{equation}
\texttt{r}_\Sigma = r_\Sigma. \label{thetaphi}
\end{equation}
The continuity of the first fundamental form is completed by requiring that the particle trajectories at the boundary be timelike.
The continuity of the angular components of the second fundamental form (extrinsic curvature) give
\begin{equation}
M = m(r_\Sigma), \label{mass}
\end{equation}
and the continuity of the remaining ($\tau$ - $\tau$) component gives
\begin{equation}
\Phi^{'}_{\Sigma} = \frac{M}{r_\Sigma ( r_\Sigma - 2M)} \label{phiprime}
\end{equation}
which, with the TOV equation, gives
\begin{equation}
p(r_\Sigma) = 0. \label{pressure}
\end{equation}
To summarize, a static spherically symmetric fluid is matched to a vacuum exterior subject to (and only to) (\ref{thetaphi}), (\ref{mass}) and (\ref{pressure}). Further restrictions are frequently imposed. In particular, if the coordinates are assumed admissible (the metric and first derivatives assumed continuous across $\Sigma$) then
\begin{equation}
e^{2 \Phi(r_\Sigma)} = 1-2 \frac{M}{r_\Sigma}, \label{admis1}
\end{equation}
and
\begin{equation}
m^{'}_\Sigma = 0 = \rho(r_\Sigma).  \label{density}
\end{equation}
Whereas (\ref{admis1}) can be achieved by a simple change in scale (of $t$ or $T$), in general, (\ref{density}) does not hold \cite{surfrho}. Condition (\ref{admis1}) is the necessary and sufficient condition for $B$ to be  continuous and continuously differentiable at $\Sigma$. Similarly, a simple change in scale makes $V$ continuous but not continuously differentiable at $\Sigma$. The wave equation (\ref{wave}) is of course invariant to these changes in scale. In summary, $B$ can be taken to be continuous and continuously differentiable at $\Sigma$, and $V$ can be taken to be continuous\cite{hevi}. 
\section{Examples}
Since the uniform density static sphere satisfies (\ref{tov1}), one might guess that all static solutions do. This is not the case. For example, the Buchdahl \cite{buchdahl} solution does not allow a region which satisfies (\ref{tov1}). In contrast, the Tolman VII solution does \cite{tolman}. (These are useful exact solutions for the study of the equation of state of neutron stars \cite{lattimer}). In what follows we demonstrate a number of physically acceptable solutions which do satisfy (\ref{tov1}). We organize the examples by way of their motivating ansatz.
\subsection{Prescribed form of $m(r)$}
The Finch-Skea solution \cite{fs} is an exact solution which gives reasonable values for the central densities of neutron stars. The  solution derives from the ansatz
\begin{equation}
m(r) = \frac{C r^3}{2(1+C r^2)},
\end{equation}
where $C$ is a constant. The line element can be given in the form
\begin{equation}
 ds^2 = v^2 dr^2 + r^2d\Omega ^2 -A^2((C_2-C_1 v)\cos(v)+(C_1+C_2 v)\sin(v)) ^2 dt^2,\label{fs}
 \end{equation}  
where  $v \equiv \sqrt{1+\omega^2}$, $\omega^2 \equiv C r^2$ and $A, C_1$ and $C_2$ are constants. Clearly $AC_2$ can be set by the scale of $t$ leaving (say) $C$ and $\beta \equiv \frac{C_{1}}{C_{2}}$ as parameters. The latter is conveniently given by
\begin{equation}
\beta = \frac{v_{\Sigma}tan(v_{\Sigma}) - 1}{tan(v_{\Sigma}) + v_{\Sigma}} \label{betav}
\end{equation} 
where $ v_{\Sigma}  \equiv \sqrt{1+\omega_{\Sigma}^2}$, or equivalently, in terms of the tenuity $\alpha \equiv \frac{r_{\Sigma}}{M}$,
\begin{equation}
\alpha = \frac{ 2 v_{\Sigma}^2}{v_{\Sigma}^2 - 1}.
\end{equation}  
 The physical restrictions 3 and 6 give, respectively, the following lower and upper bounds to $\beta$ \cite{betalimits}  
\begin{equation}
0.218 \leq \beta \leq 5.605 ,\label{betab}
\end{equation}
but the limits which follow from $B$ and $V$ are more transparently expressed in terms of $\alpha$.
Up to an irrelevant scale factor, the potential impact parameter follows immediately as, 
\begin{equation}
B(\omega)=\frac{w}{(1-\beta v)cos(v)+(\beta +v)sin(v)}. \label{finchskeaB}
\end{equation}
We find that $B$ has a local minimum for $\alpha <3$  and a local maximum with subluminal sound speed between the local maximum and minimum for $\alpha > \sim 2.768$. Some typical plots of $B$ are shown in  Fig.~\ref{Bfs}.
The full potential (up to an irrelevant scale factor) is given by
\begin{equation}
V(\omega) = \frac{l(l+1)+\frac{F(v)\omega^2}{2}-3\frac{\omega^2}{1+\omega^2}}{B^2} \label{finchskeaV}
\end{equation}
where
\begin{equation}
F(v) = \frac{2+v^2}{v^4} + \frac{1}{v^2} \frac{(\beta v+1)+(\beta-v) tan(v)}{ (\beta v-1)-(\beta+v) tan(v)}.\label{F(v)}
\end{equation}
Some typical plots of $V$ are shown in  Fig.~\ref{Vfs}. We find that there is a local minimum in $V$ (with $l = 2$) for $\alpha < \sim 2.933$ and the local minimum lies in a region with subluminal sound speed for $\alpha > \sim 2.755$ \cite{numerics}. The Finch-Skea solution therefore offers an example of a causal exact solution with trapping \cite{stability}.
\subsection{Prescribed form of $\Phi(r)$}
A class of models, some of which satisfy conditions 1 through 6, starts with the ansatz 
\begin{equation}
e^{2\Phi(r)}=D (1+E r^{2})^{n}, \label{tolmaniv}
\end{equation}
where $D$ and $E$ are constants and $n$ is an integer $ \geq 1$. The case $n=1$ is known as Tolman IV solution \cite{tolman}.  It follows immediately from (\ref{trapa}) that this solution exhibits no trapping ($B$ is monotone increasing and $V$ monotone decreasing). For $n=2$ condition 3 fails. The case $n=3$ satisfies conditions 1 through 6. It has been examined by Heintzmann \cite{Heint}, who gives the solution
\begin{equation}
  ds^2 = \frac{dr^2}{( 1-\frac{3ar^2}{2}\frac{1+C(1+4ar^2)^{-1/2} }{1+ar^2} )} + r^2d\Omega ^2-A^2( 1+ ar^2 ) ^3 dt^2.\label{Heintzmann}
\end{equation}
Again, in terms of the tenuity ($\alpha \equiv \frac{r_{\Sigma}}{M}$), we find that there is a local minimum in $V$ (with $l = 2$) for $\alpha < \sim 2.902$ and the local minimum lies in a region with subluminal sound speed for $\alpha > \sim 2.788$. A typical example is shown in Fig.~\ref{heintii} where $V$ has been matched onto the vacuum exterior at the boundary $\Sigma$, and the minimum in $V$ and sound speed limit have been indicated.
The cases $n=4$ and $n=5$ satisfy conditions 1 through 6 and have been solved by Durgapal \cite{Durg} following the formulation of Korkina \cite{korkina}. For $n=4$ the solution is given by
\begin{equation}
ds^2 = \frac{(1+Cr^2)^2dr^2}{( \frac{7-10Cr^2-C^2r^4}{7} + \frac{KCr^2}{(1+5Cr^2)^{\frac{2}{5}}})}+ r^2d\Omega ^2-A(1+Cr^2)^4 dt^2.\label{Durgapal}
\end{equation}
In this case we find that there is a local minimum in $V$ (with $l = 2$) for $\alpha < \sim 2.892$ and the local minimum lies in a region with subluminal sound speed for $\alpha > \sim 2.780$. Since $B(r)$ has a local maximum up to $\alpha \leq 3$, it is clear that the dynamical part of the potential $V$ can dominate. For $n=5$ the solution is
\begin{equation}
 ds^2 = \frac{(1+Cr^2)^3dr^2}{ (1-\frac{Cr^2(309+54Cr^2+8C^2r^4)}{112} + \frac{KCr^2}{\sqrt[3]{1+6Cr^2}})}  + r^2d\Omega ^2-A(1+Cr^2)^5 dt^2.\label{Durgapal1}
\end{equation}
We find similar results in this case. There is a local minimum in $V$ (with $l = 2$) for $\alpha < \sim 2.886$ and the local minimum lies in a region with subluminal sound speed for $\alpha > \sim 2.776$. The solutions (\ref{Heintzmann}), (\ref{Durgapal}) and (\ref{Durgapal1}) cannot represent the core region where, we note, they are acausal \cite{stability}. 
\section{Discussion}
Condition (\ref{tov1}) is proposed as the necessary condition for the internal trapping of null geodesics, and for the occurrence of resonance scattering of axial gravitational waves when the centrifugal term dominates the potential, in static spherically symmetric perfect fluids \cite{invariant}. This condition is not always satisfied. For example, it is not satisfied in the Buchdahl \cite{buchdahl} solution. We have demonstrated some physically acceptable exact solutions for which the condition is satisfied. One, the Finch-Skea solution, offers an example of a complete causal exact solution with trapping. At the very least, these examples can provide a check on numerical procedures which attempt to gauge the role that w-modes may play in real astrophysical processes. In every case studied here we have found that for resonance scattering the tenuity ($\alpha= r_{\Sigma}/M$) lies in the small range $ ~2.8 <  \alpha < ~2.9$. (In all cases, as $\alpha$ decreases, the causal boundary ($\textrm{v}_s^2=1$), if it exists,  moves out and approaches the minimum in $V$ for some minimum $\alpha$, exactly as expected.) Whereas this range is well above the Buchdahl limit of $9/4$ \cite{buchdahl1}, it is too low for, say, neutron stars \cite{lattimer}.  (Unphysical solutions with trapping and $\alpha > 5$ are known \cite{unphysical}.) It is reasonable to suggest, as has often been done, that an exact solution may reflect only part of a more realistic configuration. Boundary conditions within a distribution are easily derived from the discussion given in section \textbf{V}: $p(r)$ and $m(r)$ must be continuous to avoid surface layers (shells). In particular, $m(r)$ need not be continuously differentiable (which, at least formally, allows first-order phase transitions). All that is needed to raise $\alpha$ into a more interesting range (say $ 3<  \alpha < 10$) is the addition of an envelope. The envelope is constructed subject to the continuity of $p(r)$ and $m(r)$ at $r_{\sigma}$, where $r_{\sigma}$ is exterior to the minimum in $V$, and must allow $p(r_{\Sigma})=0$ at a finite boundary $r_{\Sigma}$ where $r_{\sigma} < r_{\Sigma}$ \cite{endnote}. 
\section*{Acknowledgments}
We thank Kostas Kokkotas, James Lattimer, Eric Poisson and Kjell Rosquist for helpful comments. This work was supported by a grant (to KL) from the Natural Sciences and Engineering Research Council of Canada.

\begin{figure}

\caption{ The potential impact parameter $B$ for the Finch - Skea solution. The curves (bottom to top) have $(\alpha, w_{\Sigma}, \beta)$ given by $(3, \sqrt{2}, 2.638)$, $(2.856, 1.529, 4)$ and $(2.768, 1.614, 6.332)$. The vertical scale is irrelevant. For $\alpha < 2.768$ the sound speed is superluminal beyond the local maximum in $B$. The curves are continued to $\Sigma$.} \label{Bfs} 
\label{fig1}
\end{figure}
\begin{figure}
\caption{ The potential $V$ (for mode number $l = 2$) for the Finch - Skea solution. The curves (bottom to top) have $(\alpha, w_{\Sigma}, \beta)$ given by $(2.933, 1.464, 3.105)$, $(2.856, 1.529, 4)$ and $(2.806, 1.575, 5)$. The vertical scale is irrelevant. The curves are continued to $\Sigma$.} \label{Vfs} 
\label{fig2}
\end{figure}
\begin{figure}
\caption{The potential $V$ (for mode number $l = 2$) for the Heintzmann solution. Here we use $\alpha = \frac{r}{M}$. $\alpha_{\Sigma} = 2.850$. $V$ has a local minimum at $\alpha = 2.717$ and the sound speed becomes superluminal below $\alpha = 2.291$. $V$ for the vacuum exterior is also shown.} \label{heintii}
\label{fig3}
\end{figure}

\begin{references}
\bibitem{C-B} Whereas at very high densities the adiabatic sound speed may not equal the
actual propagation speed of the signal, we do not distinguish between these cases here and take the conservative view that $\textrm{v}_s^2 \geq 1$ is acausal. See, for example, G.~Caporaso and K.~Brecher, Phys. Rev. D {\bf 20} 1823 (1979).
\bibitem{schwint} It is worthy of note that the standard example for w-modes, the Schwarzschild interior solution, fails essentially only at the last step. However, since the sound speed is (formally) infinite, it is physically unacceptable.
\bibitem{stability} The stability of the Finch-Skea solution (ref \cite{fs}) cannot be estimated from their Table 1. These solutions should be specified by their parameters $C$ and $\beta$ from which follow the pair $(m, r_{s})$. That is, $m$ and $r_{s}$ cannot be specified independently. In general, an exact solution can reasonably be expected to approximate only part of a full configuration. This means that the full model must be specified and a dynamical stability analysis is required for a reliable conclusion on stability.
\bibitem{units} We use geometrical units throughout. For example, the tenuity $r_{\Sigma}/M$ is actually $c^2r_{\Sigma}/GM$.
\bibitem{bschw}Relation (\ref{tov1}) returns the familiar $r<3m$ in the Schwarzschild vacuum. This field does not satisfy (\ref{reg}) since it is singular at the origin.
\bibitem{surfrho} Whereas $\rho_{\Sigma}$ is not required to be $0$, models which are otherwise physically acceptable may give physically unreasonable values of $\rho_{\Sigma}$. Uniform density models are, in this sense, the worst possible models.
\bibitem{hevi}As pointed out originally by Chandrasekhar and Ferrari \cite{CHANDRAFER}, a Heaviside discontinuity in $V$ at $\Sigma$ is of no physical consequence.
\bibitem{numerics} Although the models considered in this paper are analytic, the limits quoted on $\alpha$ are the result of detailed numerical studies.
\bibitem{betalimits}The upper limit differs from the value given in \cite{fs} and may be traced to an error in their equation (6.14). It is worth noting that the sound speed fails to be monotone decreasing with $r$ for $\alpha \geq 2.679$. The sound speed is subluminal throughout the solution for $\alpha > 2.786$.
\bibitem{invariant} The quantities which enter (\ref{tov1}) can be invariantly defined.
\bibitem{endnote} It is worth noting that if a solution  represents only part of a more realistic configuration, then causality, but not stability, conditions need be applied to the solution.
\end{references}
\end{document}